\documentclass[9pt,twocolumn,superscriptaddress,floatfix,pra]{revtex4-1}

\usepackage{amsthm}
\usepackage{amsmath}
\usepackage{bbm, dsfont}
\usepackage{graphicx}
\usepackage[colorlinks=true,citecolor=blue,urlcolor=black,linkcolor=blue]{hyperref}
\usepackage[table]{xcolor}
\usepackage{booktabs}

\newcommand{\be}{\begin{equation}}
\newcommand{\ee}{\end{equation}}

\newcommand{\del}[1]{}

\begin{document}
\title{High-detection efficiency and low-timing jitter with amorphous superconducting nanowire single-photon detectors} 
\author{Misael Caloz}\email{misael.caloz@unige.ch}
\affiliation{Group of Applied Physics, University of Geneva, CH-1211 Geneva, Switzerland}
\author{Matthieu Perrenoud}
\affiliation{Group of Applied Physics, University of Geneva, CH-1211 Geneva, Switzerland}
\author{Claire Autebert}
\affiliation{Group of Applied Physics, University of Geneva, CH-1211 Geneva, Switzerland}
\author{Boris Korzh}
\altaffiliation[]{Now at Jet Propulsion Laboratory, Pasadena, CA 91109 USA}
\affiliation{Group of Applied Physics, University of Geneva, CH-1211 Geneva, Switzerland}
\author{Markus Weiss}
\author{Christian Sch\"onenberger}
\affiliation{Department of Physics, University of Basel, CH-4056 Basel, Switzerland}
\author{Richard J.~Warburton}
\affiliation{Department of Physics, University of Basel, CH-4056 Basel, Switzerland}
\author{Hugo Zbinden}
\affiliation{Group of Applied Physics, University of Geneva, CH-1211 Geneva, Switzerland}
\author{F\'elix Bussi\`eres} 
\affiliation{Group of Applied Physics, University of Geneva, CH-1211 Geneva, Switzerland}
\affiliation{ID Quantique SA, CH-1227 Carouge, Switzerland}


\begin{abstract}

Recent progress in the development of superconducting nanowire single-photon detectors (SNSPDs) made of amorphous material has delivered excellent performances, and has had a great impact on a range of research fields. Despite showing the highest system detection efficiency (SDE) ever reported with SNSPDs, amorphous materials typically lead to lower critical currents, which impacts on their jitter performance. Combining a very low jitter and a high SDE remains a challenge. Here, we report on highly efficient superconducting nanowire single-photon detectors based on amorphous MoSi, combining system jitters as low as 26~ps and a SDE of 80\% at 1550~nm. We also report detailed observations on the jitter behaviour, which hints at intrinsic limitations and leads to practical implications for SNSPD performance. 

\end{abstract}
\maketitle

Since their first demonstration, superconducting nanowire single-photon detectors (SNSPDs) have emerged as a key technology for optical quantum information processing~\cite{Goltsman2001a, Hadfield2009}. Their low dark count rate, fast response time, small jitter, and high efficiency favours their use in various demanding quantum optics applications such as quantum key distribution~\cite{Takesue2007}, quantum networking~\cite{Bussieres2014}, device-independent quantum information processing~\cite{Shalm2015a}, deep-space optical communication~\cite{Shaw14} and IR-imaging~\cite{Allman2015a,Zhao2017}. Notably, SNSPDs can be integrated in photonic circuits~\cite{Sprengers2011,Rath2015}. 


One recent advance in the SNSPD field has been the introduction of amorphous superconductors such as tungsten silicide (WSi)~\cite{Marsili2013}, molybdenum silicide (MoSi)~\cite{korneeva2014,Verma2015,Caloz2017a} and molybdenum germanium (MoGe)~\cite{Verma2014b}. SNSPDs based on these materials currently have the highest reported system detection efficiencies (SDE) (93\% for WSi~\cite{Marsili2013}), as well as a high fabrication yield~\cite{Allman2015a}. 


The jitter of an SNSPD denotes the timing variation of the arrival time of the detection pulses. The jitter by itself is a crucial characteristic for time-resolved measurements such as light detection and ranging,  high-speed quantum communication, and lifetime measurement of single-photon sources. Typically, for a Gaussian distribution, the jitter is quantified using the full width at half maximum (FWHM) of the distribution. Despite showing the highest SDE ever reported with SNSPDs, amorphous materials operate at low bias currents and hence showed until now a time jitter rather high compared to what can be achieved with NbN~\cite{Wu17,You2013,Shcheslavskiy2016} and NbTiN~\cite{Zadeh2017}. A wide range of values have been reported for different geometries and materials, typically from tens to hundreds of picoseconds. Some recently reported values range between $\sim$15~ps (NbN\cite{Wu17}, NbTiN~\cite{Zadeh2017}), $\sim$18~ps (NbN\cite{You2013,Shcheslavskiy2016}), and 76~ps for amorphous material (MoSi)~\cite{Verma2015}.


In this work, we report on our results on the low timing jitter and high SDE of our MoSi SNSPDs. We measured the system jitters and SDE for several devices and obtained jitters (FWHM) as low as 26~ps and saturated SDE of 80~\% or more at telecom wavelength. We also report on detailed observations on the jitter behaviour, which hints at intrinsic limitations and leads to practical implications for SNSPD performance. 


The SNSPDs are fabricated out of a 7~nm-thick film of amorphous Mo$_{0.8}$Si$_{0.2}$ deposited by co-sputtering with a DC and RF bias on the molybdenum and silicon targets, respectively. X-ray diffraction measurements have been performed, confirming the amorphous nature of the MoSi. The fabrication is done in the following way: (i) a metallic mirror is evaporated on a thermally oxidised silicon wafer, (ii) a silicon dioxide (SiO$_2$) layer with a $\sim \lambda$/4 thickness is deposited by RF sputtering, (iii) the MoSi film is deposited, capped with a 3~nm amorphous silicon (a-Si) layer and covered by $\sim 50$~nm of SiO$_2$. By choosing correctly the thickness of the two SiO$_2$ layers, constructive interference inside the structure maximises the absorption in the MoSi layer~\cite{Anant2008}. The film is patterned as a meandered wire covering a total surface area of 16$\times$16~$\mathrm{\mu}$m$^2$ by a combination of e-beam lithography and reactive ion etching. One wafer contains devices with different widths (100-180~nm) and fill factors (fraction of active area). A self-aligning technique is used to ensure optimal coupling to the optical fibre~\cite{Miller2011}. The room temperature resistance of our devices is a few $M\Omega$, depending on the geometry of the nanowire and of the meander. The current density at $I_{sat}$ is typically around 3~MA/cm$^2$ and is similar for all devices, more details can be found in the Supplementary Material.
 
\begin{figure}[t!] 
	\vspace{0.5cm}
	\includegraphics[width=85mm]{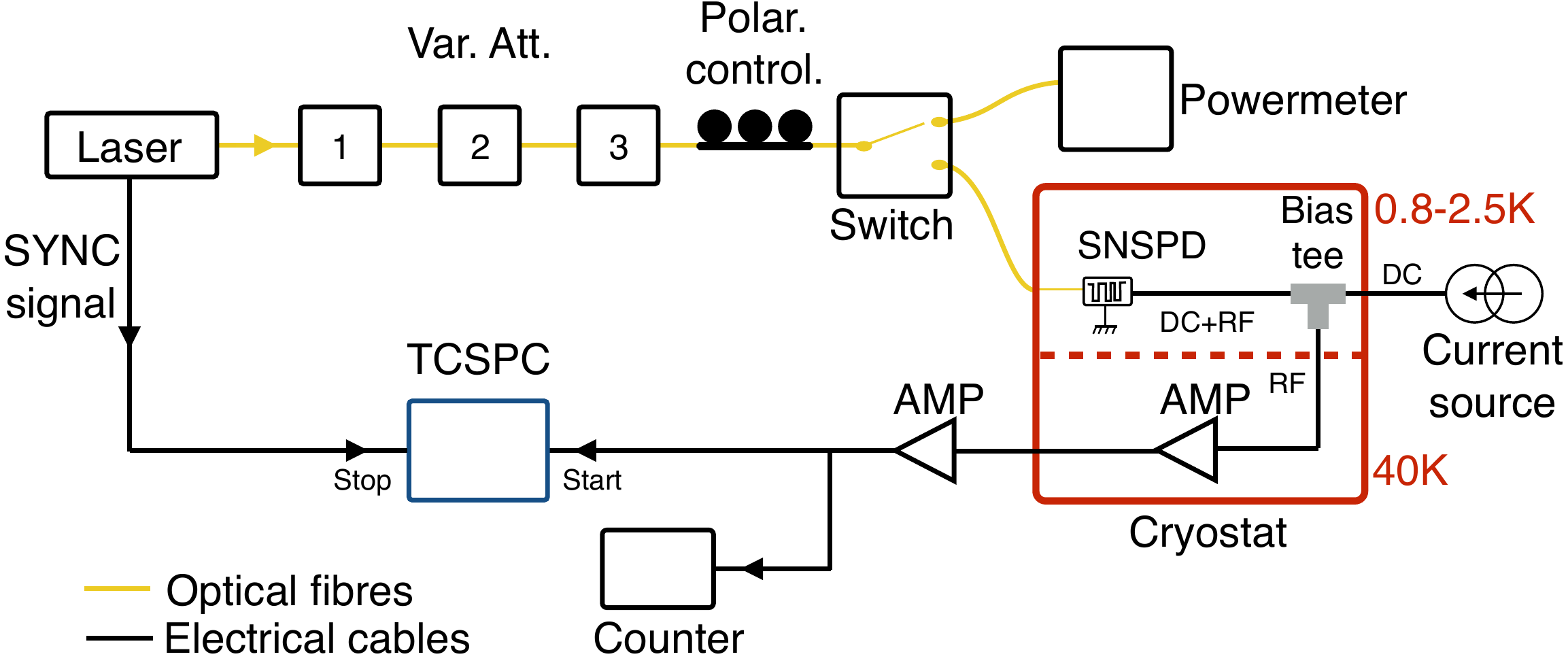}
	\caption{Schematic view of the setup for measuring both system jitter and the efficiency of the SNSPDs. For jitter and SDE measurement, the counter and TCSPC modules were not used, respectively.}\label{fig1}
\end{figure}

The detectors are mounted in a sorption cryostat reaching 0.8~K. For measuring the jitter of the SNSPDs, a TCSPC module (Becker \& Hickl, SPC-130) with a constant fraction discriminator (CFD) was set up and a 6~ps (FWHM) pulse width fibre laser (Nuphoton Technologies) at 1560~nm was used as the source, as shown in Fig.~\ref{fig1}. The power of the source was attenuated to the single photon level by variable attenuators. The single-photon-response voltage pulse is amplified by a custom low-noise amplifier cooled to 40~K and by a secondary amplifier at room temperature. The cryogenic preamplifier is not necessary to operate the detectors but it does provide a larger signal-to-noise ratio (SNR). The pulse
	polarity has no impact on the measured detector performances. For SNSPDs, the distribution of the intervals between the ``Start" and the ``Stop" signals typically show a Gaussian profile, from which the system jitter can be extracted. The CFD of the TCSPC module ensures that the discrimination of the electrical pulse of the detector is done relative to its amplitude. For measuring the SDE we used a continuous wave (CW) polarized laser at 1550~nm attenuated down to $10^5$ photons/second by three variables attenuators in series and a calibrated powermeter, see the Supplementary Material for more details. The input light polarization was set to optimize the number of counts of the SNSPDs. Fig.~\ref{fig1} shows the schematic view for both jitter and SDE measurements. The measured jitter of the TCSPC module itself is 9~ps. We confirmed that our devices do not suffer from after pulsing by using a setup with a time to digital converter and a pulsed laser.

\begin{figure}[t!] 
	\includegraphics[width=85mm]{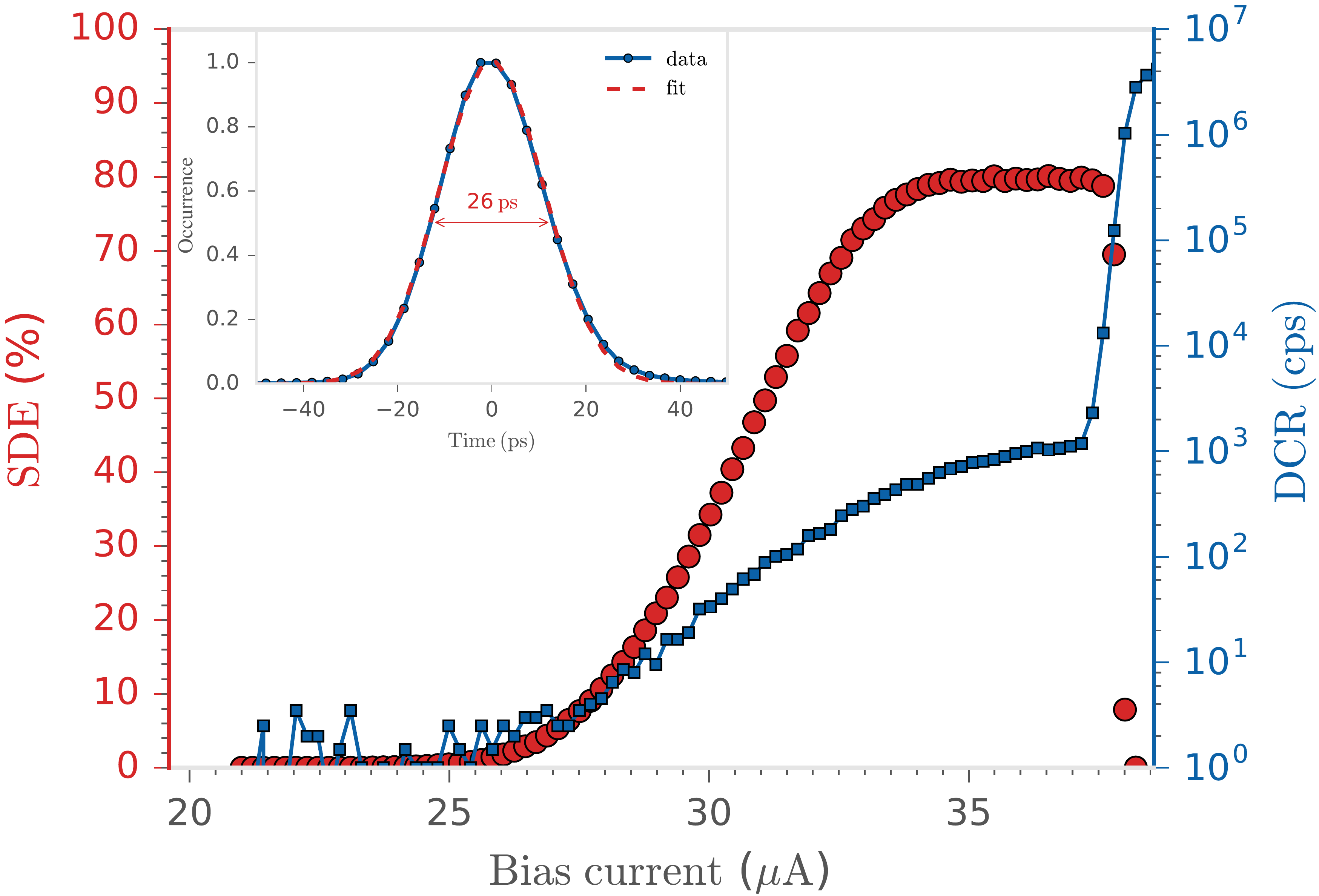}
	\caption{System detection efficiency (red circles) and dark count rate (blue squares) as a function of the bias current for device \#5, at 1550~nm and 0.8~K. Error bars are too small to be seen. Inset:~System jitter for the same device at $\mathrm{I_{b}} = 37$~$\mu$A, the blue and red lines indicate the data and the gaussian fit, respectively. The system jitter measured is 26~ps (FWHM) and is indicated by the double arrow.}\label{fig2}
\end{figure}

The SNSPD devices that we tested all have critical current above 30 uA, which results in detection pulses with large amplitudes. This greatly reduces the jitter component due to the noise, allowing us to reach very low jitters while keeping high efficiencies. We measured the system jitters and the SDE for tens of devices. At the operating temperature of 0.8~K and for 1550~nm, all tested devices exhibited a plateau region and very similar performances according to their designs, all of them showed SDE $>$~74\% and system jitters $<$~45~ps at the same time, selected devices for this paper are shown in Tab.~\ref{tab1}. In particular, we obtained a device combining a system jitter as low as 26~ps (FWHM) for a SDE of 80.1~\%~$\pm$~0.9~\% as shown in Fig.~\ref{fig2}, and another one combining a SDE of 85.8~\%~$\pm$~0.9~\% and system jitter of 44~ps. The DCR of $\leq$1000~cps, mainly due to the black body radiation, can be significantly reduced by installing fibre based filters. The uncertainty on the efficiency measurement has been estimated by an error propagation calculation, details on the computation are explained in the Supplementary Material. 
	\begin{table}
		\caption{List of selected devices with their characteristics.}
		\label{tab1}
		\begin{ruledtabular}
			\begin{tabular}{lcccr}
				Detector & width (nm) & fill factor & SDE (\%) & Jitter (ps) \\
				\hline
				\#1		& 150 & 0.7 & 85.8 & 44.2 \\ 
				\#2		& 150 & 0.7 & 82.3 & 35.4 \\ 
				\#3  	& 160 & 0.6 & 80.2 & 32.7 \\ 
			    \#4     & 150 & 0.6 & 76.5 & 30.1 \\ 
				\#5 	& 160 & 0.5 & 80.1& 26.1 \\ 
				\#6 	& 150 & 0.5 & 74.6 & 28.6 \\ 
			\end{tabular}
		\end{ruledtabular}
	\end{table}

The measured system timing jitter $j_\mathrm{sys}$ can be decomposed into three main parts: (i) a noise component coming from the electronic readout noise, (ii) a setup component from laser pulse width and TCSPC module, and (iii) a component which is intrinsic to the detection process (hotspot dynamic and geometric effect~\cite{Zhao2017}). Improving the detector pulse amplitude has significantly decreased the noise-induced jitter component, allowing us to observe intrinsic jitter behaviour which was not accessible until now with amorphous materials. While the two first components are well-known contributions, it remains unclear how the intrinsic jitter contributes to $j_\mathrm{sys}$~\cite{You2013,Calandri2016}. The spread of the reported system jitter values in the literature makes it difficult to determine the origin of the intrinsic jitter of a device quantitatively, and the mechanism of this intrinsic jitter is still not completely understood~\cite{Zhao2011,You2013,Calandri2016,Connor2011}. By analysing the bias current dependence of the system jitter for several devices, we can extract the contribution of the intrinsic jitter and reveal its behaviour as the detectors efficiency reaches saturation. Assuming the noise ($j_\mathrm{noise}$), intrinsic ($j_\mathrm{int}$) and setup ($j_\mathrm{setup}$) contributions to the system jitter are independent~\cite{Calandri2016,You2013}, we can write the system jitter as
\begin{equation}
j_\mathrm{sys} = \sqrt{j_\mathrm{noise}^2 + j_\mathrm{setup}^2 + j_\mathrm{int}^2},
\label{eq1}
\end{equation}
where the intrinsic jitter itself is a combination of the jitter coming from the hotspot dynamics and the geometric effects, \\ $j_\mathrm{int} = \sqrt{j_\mathrm{hotspot}^2 + j_\mathrm{geometric}^2}$. Here, the $j_\mathrm{hotspot}$ and $j_\mathrm{geometric}$ cannot be estimated independently. Nevertheless, the intrinsic jitter $j_\mathrm{int}$ can be estimated if the other contributions are known: $j_\mathrm{setup}$ is given by the laser specification sheets and by the TCSPC module measurement, while the noise-induced jitter ($j_\mathrm{noise}$) was estimated from
\begin{equation}
j_\mathrm{noise} = 2 \sqrt{2 \ln(2)}\ \frac{\sigma_\mathrm{RMS}}{SR},
\label{eq2}
\end{equation}
where $\sigma_\mathrm{RMS}$ is the RMS value of the electronic noise, and $SR$ the slew rate of the electrical pulse coming from a detection event in the SNSPD, both measured on an oscilloscope having a 6~GHz bandwidth, more details can be found in the Supplementary Material.

Figure~\ref{fig3a} shows the evolution of the system jitter as a function of the bias current for different devices listed in Tab.~\ref{tab1}. In order to compare them, the bias current is normalized to the saturation current ($\mathrm{I_{sat}}$), which we defined as the bias current at which the SDE reaches 90\% of its maximum value at the plateau. The jitter value for devices \#1 and \#2 are higher than for the other ones. These devices have a higher fill-factor and are also longer. Their larger jitter could possibly be attributed to a larger geometric effect, although this cannot be confirmed from these measurements alone. We plotted the different system jitter components using Eq.~(\ref{eq1}) and (\ref{eq2}) for device \#4 in Fig.~\ref{fig3b}.

\begin{figure}[] 
	\includegraphics[width=85mm]{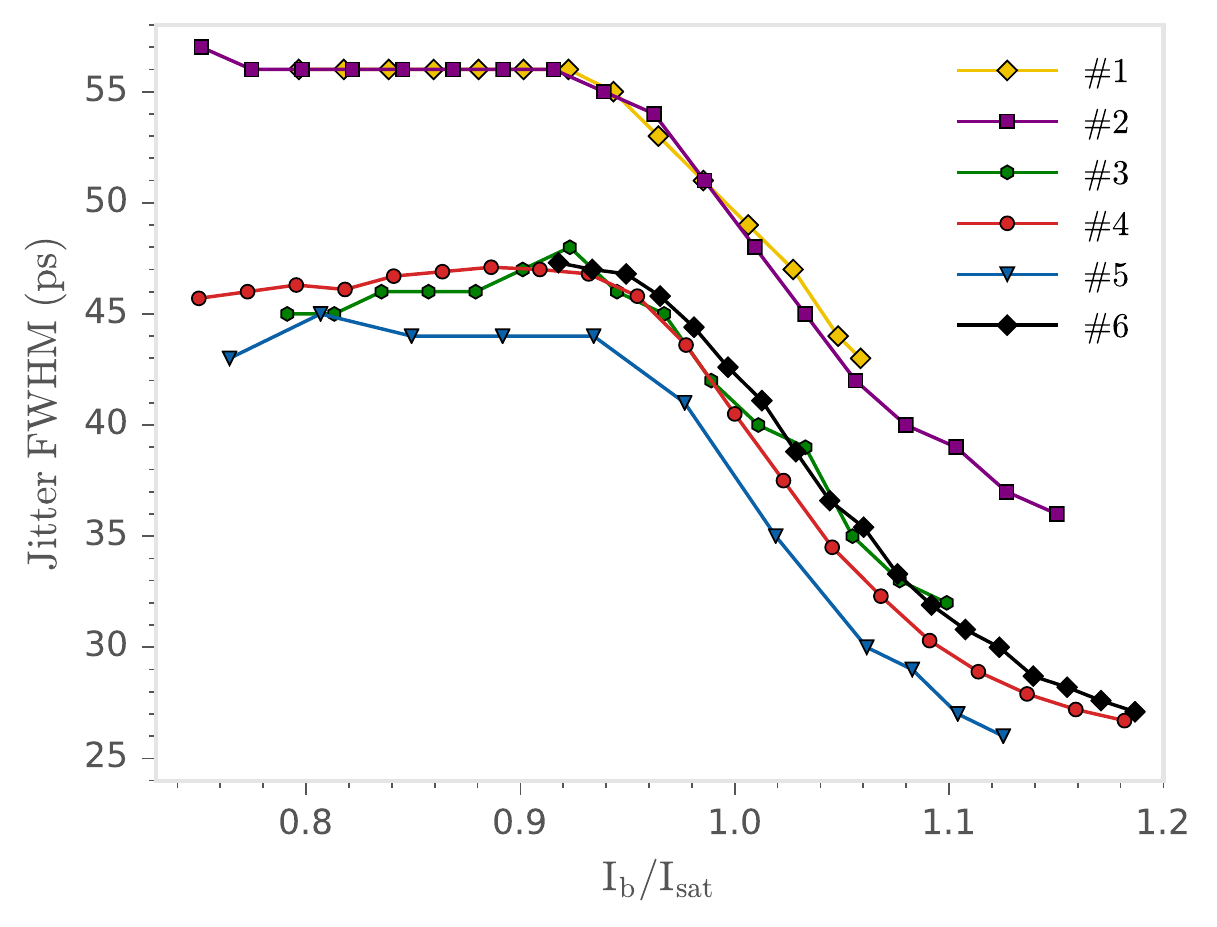}
	\caption{Jitter (FWHM) as a function of $\mathrm{I_{b}}$ normalized to the saturation current ($\mathrm{I_{sat}}$) for different devices shown in Tab.~\ref{tab1}. Here, $\mathrm{I_{sat}}$ is defined as the bias current at which the SDE reaches 90\% of its maximum. Error bars are too small to be seen.}\label{fig3a}
\end{figure}

\begin{figure}[] 
	\includegraphics[width=85mm]{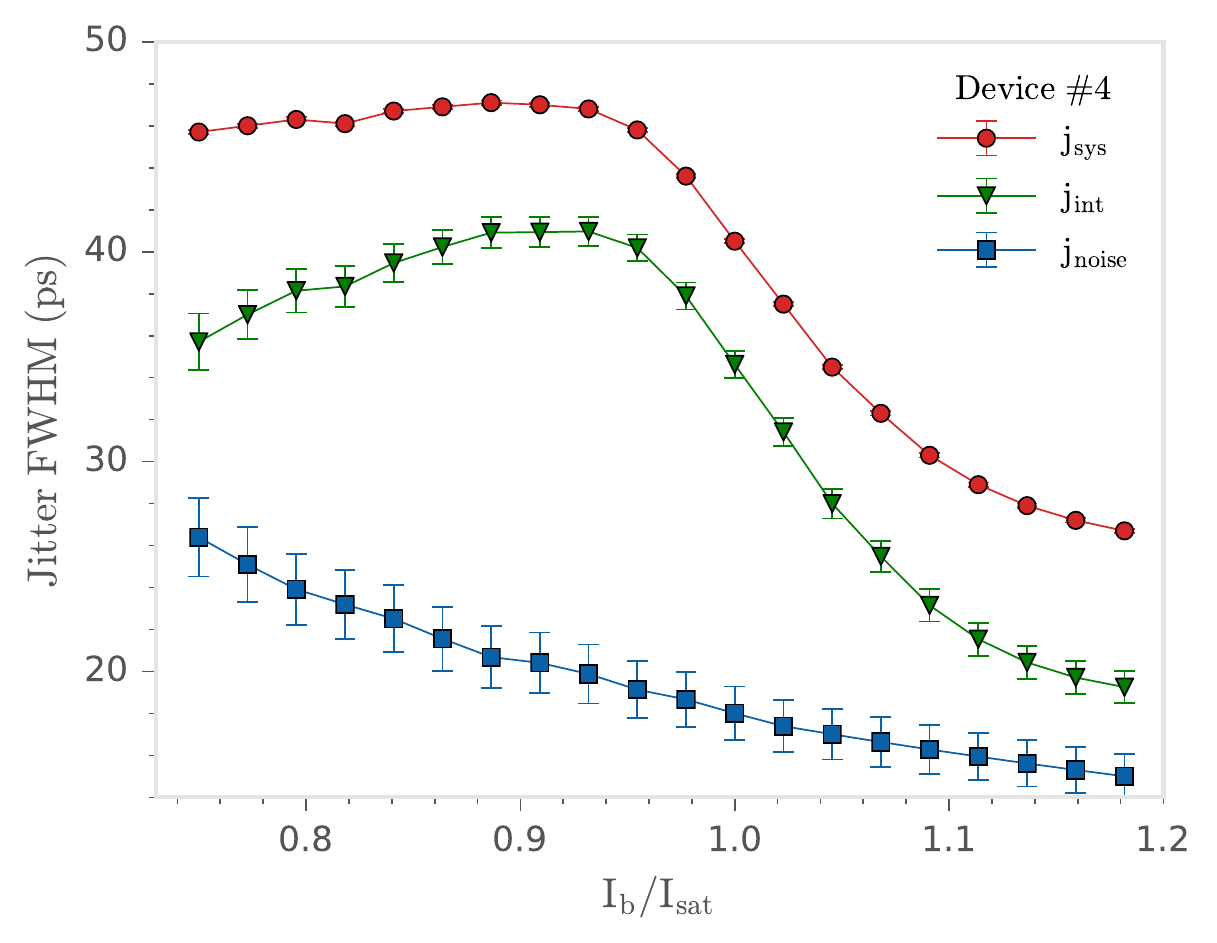}
	\caption{Different jitter components (FWHM) as a function of $\mathrm{I_{b}}$ normalized to the saturation current ($\mathrm{I_{sat}}$) for device \#4 with their error bars. The coloured lines represent the different jitter components in the following way, red: measured system jitter, blue: estimated noise-induced jitter using Eq.~(\ref{eq2}), green: computed intrinsic jitter using Eq.~(\ref{eq1}).}\label{fig3b}
\end{figure}

For high bias currents, the noise-induced jitter becomes very small, an improvement in the amplification chain could possibly reduce it even more~\cite{Calandri2016,Shcheslavskiy2016}, which could potentially lead to system jitters below 20~ps. We note that the intrinsic jitter $j_\mathrm{int}$ strongly depends on the applied bias current. 
From Fig.~\ref{fig3a} and Fig.~\ref{fig3b}, for all devices (with different widths, lengths and fill factors) the following points can be highlighted: i) $j_\mathrm{sys}$ is constant for low bias currents, ii) $j_\mathrm{sys}$ exhibit the same inflexion point close to $\sim 0.92 \, \mathrm{I_{sat}}$, iii) by increasing the bias current above the inflexion point, the system and intrinsic jitters decrease significantly, iv) the jitter flattens close to $\sim 1.2 \, \mathrm{I_{sat}}$ and could potentially reach an optimal value. These observations are relevant for studying the detection mechanism in SNSPDs~\cite{Sidorova2017} but this analysis is beyond the scope of this study and is left for future work. Points iii) and iv) have implications for SNSPDs performances, namely that operation well into the plateau ($\mathrm{I_{b}} > \mathrm{I_{sat}}$) is necessary to reach an optimal jitter value.

\begin{figure}[t!] 
	\includegraphics[width=85mm]{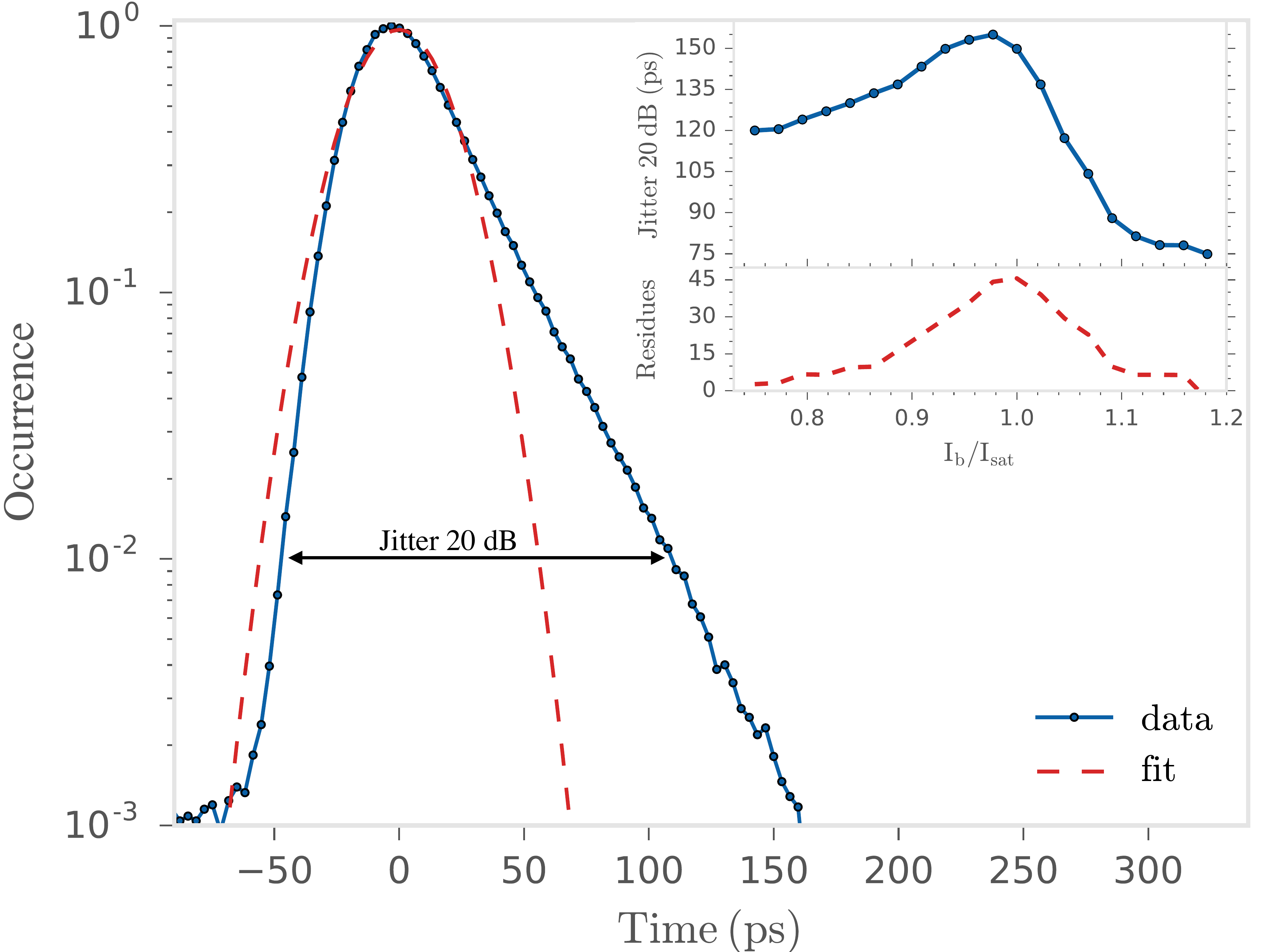}
	\caption{System jitter distribution on a logarithmic scale at a bias current equal to $\mathrm{I_{sat}}$. The blue and red lines represent the data and the gaussian fit, respectively. The double arrow indicates where the system jitter at $-20$~dB $j_\mathrm{sys}\mathrm{(-20\,dB)}$ is extracted. Inset: $j_\mathrm{sys}\mathrm{(-20\,dB)}$ and its residues from what is expected with a gaussian distribution ($\mathrm{residues} = j_\mathrm{sys}\mathrm{(-20\,dB)} - j_\mathrm{gauss}\mathrm{(-20\,dB)}$). Error bars are too small to be seen.}
	\label{fig4}
\end{figure}

Interestingly, the jitter histogram of all tested devices is asymmetric and non-gaussian in the vicinity of $\mathrm{I_{sat}}$.   
Figure~\ref{fig4} shows such a distribution measured at $\mathrm{I_{b}} = \mathrm{I_{sat}}$. The asymmetry  consists of a long exponentially decaying tail after the maxima of the histogram. This is the ``transition'' region between the ``probabilistic'' regime, where the absorption of a photon leads to a resistive region with a small probability, and the ``deterministic'' regime (the plateau), where photon absorption leads to a resistive region with almost certainty. The asymmetry however mostly disappears outside of the transition region, where it tends to be much more gaussian. The same observations have recently been reported and discussed in a theoretical framework to understand better the detection mechanism in SNSPDs.~\cite{Sidorova2017}. The first inset of Fig.~\ref{fig4} shows the system jitter at 20~dB $j_\mathrm{sys}\mathrm{(-20\,dB)}$ below the maxima of the histogram. 
To highlight the non-gaussian behaviour, the residues between the $j_\mathrm{sys}\mathrm{(-20\,dB)}$ and the gaussian distribution is shown on the second inset. Given that the setup ($j_\mathrm{setup}$) and noise ($j_\mathrm{noise}$) jitter distributions are gaussian, this evolution of the asymmetry can only be explained by an intrinsic contribution. From an application point of view, it is clear here too that the optimal SNSPD operation ($j_\mathrm{sys}\mathrm{(FWHM)}$ and $j_\mathrm{sys}\mathrm{(-20\,dB)}$) is reached but also when the bias current is greater than $\sim 1.1 \, \mathrm{I_{sat}}$. This means again that a detector with a very large deterministic region will show intrinsically better performances in term of both $j_\mathrm{sys}\mathrm{(-20\,dB)}$ and $j_\mathrm{sys}\mathrm{(FWHM)}$. This point is particularly relevant for applications where a low $j_\mathrm{sys}\mathrm{(-20\,dB)}$ is mandatory~\cite{Amri2016}, such as quantum key distribution and time-resolved measurements, where the visibility of a Bell state measurement on photonic qubits created  at random times will be directly affected by the ability of the detectors to resolve the arrival time of the photons~\cite{Bussieres2014}.

In conclusion, we reported on highly efficient superconducting nanowire single-photon detectors based on amorphous MoSi operating at 0.8~K combining a system jitter as low as 26~ps and a SDE greater than 80\% at 1550~nm at the same time. We achieved high bias currents, and we showed that the timing jitter is limited by noise and by an intrinsic component. The observations of its behaviour indicate that the system jitter might reach an optimal value for a high bias current values, hinting at an intrinsic limit. A non-gaussian tail increasing the system jitter at $-$20~dB has also been observed and quantified, having direct implications for applications such as quantum key distribution where low jitters are crucial. Our results, and in particular the fact that we can study the jitter behaviour well into the plateau, could lead to insights in the study of the detection mechanism in SNSPDs~\cite{Engel2015,Sidorova2017}. In this work, we could not isolate the contribution of the geometric jitter from the one due to hotspot dynamics. This could be attempted by either using a double-ended readout amplifier~\cite{Calandri2016}, or by using detectors made of a very short wire. Such studies are left for future work. 



The authors would like to acknowledge Nuala Timoney, Jelmer J.~Renema and Varun B.~Verma for useful discussions, Claudio Barreiro and Daniel Sacker for technical assistance, the Swiss NCCR QSIT (National Center of Competence in Research - Quantum Science and Technology) and the Swiss CTI (Commission pour la Technologie et l'Innovation) for financial support.



%


\newpage \null \newpage
\onecolumngrid
\appendix




\section*{Estimation of the uncertainty on the system detection efficiency}
\subsection*{Introduction}
\begin{figure}[h!] 
	\includegraphics[width=.7\textwidth]{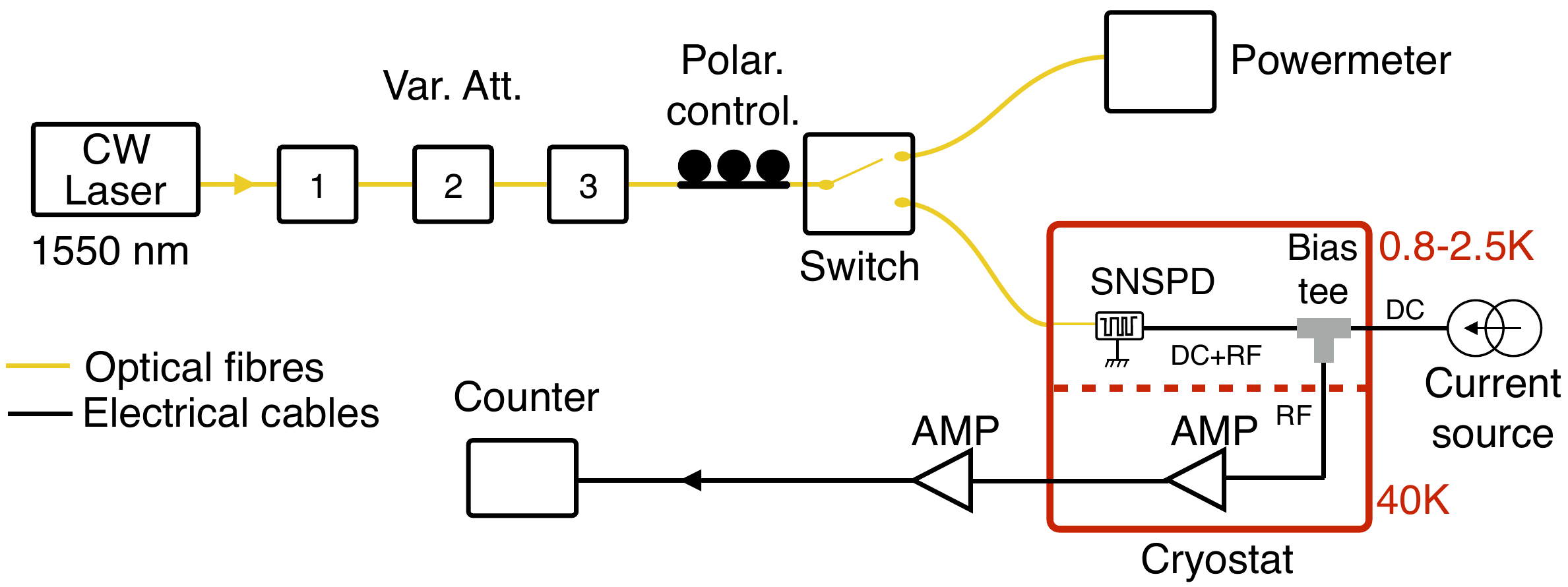}
	\caption{Schematic view of the setup for measuring the system detection efficiency of the SNSPDs.}
	\label{fig1supp}
\end{figure}

In this section, we describe the details of the measurements of the system detection efficiency (SDE) and the computation of its uncertainty using error propagation. The SDE measurement relies on the calibration of many components, but most importantly on a powermeter, calibrated by METAS (Swiss federal institute of metrology). The schematic view of the setup is shown in Fig.~\ref{fig1supp}, and all parameters with their respective uncertainty are described in Tab.~\ref{tab1:supp}.

\begin{table}[h!]
	\caption{List of all parameters, their signification and uncertainty.}
	\label{tab1:supp}
	\begin{ruledtabular}
		\begin{tabular}{llll}
			Parameter & Signification & Uncertainty & Comments  \\
			\hline
			$\eta$			  & System detection efficiency (SDE) 							               & $\sigma_\eta$ & Estimated from error propagation\\ 
			$N_\gamma$	& Number of photons per second arriving onto the detector			      & $\sigma_{N_{\gamma}}$ & Estimated from error propagation \\ 
			$PCR$			& Photon count rate														              & $\sigma_{PCR}$ & Measured\\ 
			$DCR$			& Dark count rate															              & $\sigma_{DCR}$ & Measured\\ 
			$R_{switch}$ & Ratio between port 1 and port 2 of the optical switch               & $\sigma_{R_{switch}}$ & Measured \\	
			$R_{i,att}$       & Attenuation ratio of the i-th attenuator                                    & $\sigma_{R_{att}}$ & Measured \\
			$P_M$			& Reference power measurement                                                & $\sigma_{P_M}$ & Calibrated by METAS \\
			$P_{i,att}$		& Power measurement with the i-th attenuator set ON                 & - & Calibrated by METAS \\ 
			$CF$			 & Correction factor on the power measurement				            & - & Calibrated by METAS \\ 
			$NLF_{high}$ & Non-linearity correction factor for high power		                  & - & Calibrated by METAS \\ 
			$NLF_{low}$	 & Non-linearity correction factor for low power				          & - & Calibrated by METAS \\ 
			$E_\gamma$	& Photon energy																             & - & Negligible\\  	
			$R_{pc}$		& End-face reflection coefficient of optical fibre    		                & -  & Negligible\\ 
		\end{tabular}
	\end{ruledtabular}
\end{table}
The SDE is given by the following formula:
\begin{equation}
	\label{eqs1}
	\eta~(\%) = \frac{PCR - DCR}{N_\gamma}
\end{equation}
where:
\begin{equation}
	\label{eqs2}
	N_\gamma = \frac{P_M \cdot R_{switch}}{1-R_{pc}} \times \frac{1}{E_\gamma} \times \frac{P_{1,att}}{P_{M}} \times \frac{P_{2,att}}{P_{M}} \times \frac{P_{3,att}}{P_{M}} \times \frac{1}{CF \cdot NLF_{high}} \times \left(\frac{NLF_{high}}{NLF_{low}}\right)^3
\end{equation}

\subsection*{Error propagation}
Given Eq.~\ref{eqs1} and \ref{eqs2}, the error propagation gives:
\begin{equation}
	\label{eqs3}
	\left(\frac{\sigma_\eta}{\eta}\right)^2 =   \left( \frac{\sqrt{\sigma_{PCR}^2 + \sigma_{DCR}^2}}{PCR - DCR}\right)^2       +    \left(       \frac{\sigma_{N_{\gamma}}}{N_{\gamma}}          \right)^2
\end{equation}
where:
\begin{equation}
	\label{eqs4}
	\left(       \frac{\sigma_{N_{\gamma}}}{N_{\gamma}}          \right)^2 =  \left(\frac{\sigma_{P_M}}{P_M}\right)^2 + \left(\frac{\sigma_{R_{switch}}}{R_{switch}}\right)^2 + 3 \cdot \left(\frac{\sigma_{R_{att}}}{R_{att}}\right)^2
\end{equation}
$\frac{\sigma_{R_{att}}}{R_{att}}$ is the relative uncertainty on the ratio $\frac{P_{i,att}}{P_{M}}$. $R_{pc}$, $CF$, $NLF_{high}$, and $NLF_{low}$ have negligible contribution to the SDE uncertainty. The uncertainty on $E_{\gamma}$ is directly related to the line width of the laser source, which is also negligible. 

\subsection*{Photon count rate and dark count rate}
\begin{figure}[t!] 
	\includegraphics[width=85mm]{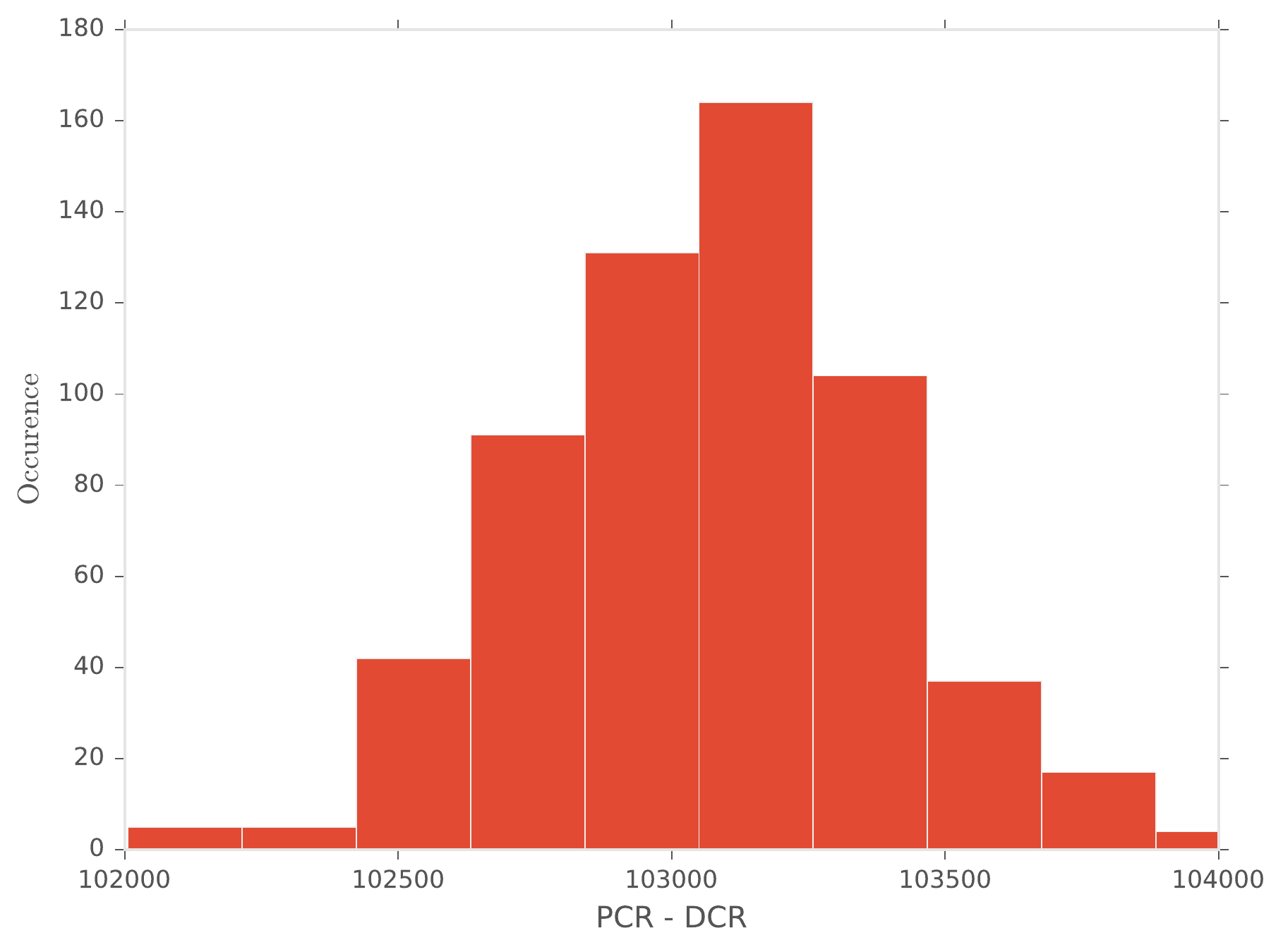}
	\caption{Typical SNSPD counts distribution, within a time interval that correspond to a typical SDE measurement ($\sim$~10~minutes), with the same light power and polarization optimization. The standard deviation of this distribution is $\sigma_{PCR-DCR} = 344.2$.}
	\label{fig2supp}
\end{figure}

$\frac{\sqrt{\sigma_{PCR}^2 + \sigma_{DCR}^2}}{PCR - DCR}$ is the relative uncertainty on the photon count rate minus the dark count rate. It has been measured 6 different times within a time interval that correspond to a typical SDE measurement ($\sim$ 10 minutes), with the same light power and polarization optimization. $\sigma_{PCR}$ and $\sigma_{DCR}$ has been extracted from these distributions. One of them is shown in Fig.~\ref{fig2supp}. The measured $\sigma_{PCR}$ and $\sigma_{DCR}$ include all sources of uncertainty due to fluctuations that can happen during a measurement, which mean:
\begin{itemize}
	\item input light polarization stability
	\item laser stability
	\item laser intensity noise 
	\item the attenuation stability
	\item ratio switch stability
\end{itemize}
The standard deviation value of the distribution shown in Fig.~\ref{fig2supp} ($\sigma_{PCR-DCR} = 344.2$) is very close to the shot noise of the laser, meaning that the other components listed above have a very small contribution to the SDE uncertainty. 
During a typical SDE measurement, the PCR and DCR are counted for different bias current values. Because we have a plateau region (assumed to be flat), where $PCR - DCR$ is constant we average the SDE over different points,typically $\sim 10$ (see Fig.2 of the manuscript). Finally we have: $\frac{\sqrt{\sigma_{PCR}^2 + \sigma_{DCR}^2}}{PCR - DCR}  / \sqrt{10}= 0.14 ~\%$. 

\subsection*{Power measurement}
$\frac{\sigma_{P_M}}{P_M}$ is the relative uncertainty on power measurement of our calibrated powermeter. It has been calibrated by METAS. $\frac{\sigma_{P_M}}{P_M} =  0.70~\%$.
\subsection*{Switch ratio}
$\frac{\sigma_{R_{switch}}}{R_{switch}}$ is the relative uncertainty on the switch ratio ($R_{switch} = \frac{P_1}{P_2}$), it is the main source of uncertainty in our setup. It has been measured more than 15 times with different setup configuration (input light polarization, laser power, different days, after plugging/unplugging the fibre, etc...). $P_1$ and $P_2$ are measured with the same calibrated powermeter, thus the uncertainty on the calibration do not propagate for $\frac{P_1}{P_2}$. This measurement includes also the repeatability of the optical switch ratio. We also note that the precision of the powermeter reading (number of digits) is high enough to not affect the computation of $\sigma_{R_{switch}}$. $\frac{\sigma_{R_{switch}}}{R_{switch}} = 0.80~\%$.

\subsection*{Attenuation repeatability, uncertainty}
$\frac{\sigma_{R_{att}}}{R_{att}}$ is the relative uncertainty on the ratio $\frac{P_{i,att}}{P_{M}}$. The uncertainty on $P_{i,att}$ and $P_{M}$ come from the absolute calibration of the powermeter, thus the uncertainty on the calibration does not propagate for the three ratios $\frac{P_{i,att}}{P_{M}}$.  The only contribution to $\frac{\sigma_{R_{att}}}{R_{att}}$ is the repeatability of the three attenuators. The uncertainty of $\frac{P_{i,att}}{P_{M}}$ has been measured with one attenuator, assuming that it is the same for the three attenuators. Similarly to the switch ratio $R_{switch}$, we note that the precision of the powermeter reading is high enough to not affect the computation of $\sigma_{R_{att}}$. By measuring enough values, we calculated the standard deviation of the distribution. $\frac{\sigma_{R_{att}}}{R_{att}} = 0.07~\%$.

\subsection*{Summary}
Tab.~\ref{tab2:supp} summarizes the different contributions and their respective uncertainty. 
\begin{table}[h!]
	\caption{Relative uncertainties of the different parameters and SDE.}
	\label{tab2:supp}
	\begin{ruledtabular}
		\begin{tabular}{lll}
			Source & Parameter & Relative uncertainty (\%) \\
			\hline
			PCR - DCR 	& $PCR - DCR$					& 0.14 \\
			Power measurement & $P_M$		& 0.70 \\
			Ratio switch 			& $R_{switch}$		& 0.80 \\
			Attenuation repeatability & $R_{att}$ & 0.07 \\
			\hline
			SDE 					& $\eta$			&  \textbf{1.08} \\
			
		\end{tabular}
	\end{ruledtabular}
\end{table}

\section*{Noise and setup jitter}
\subsection*{Noise jitter}
Eq.~2 of the manuscript is reminded below:

\begin{equation}
	j_\mathrm{noise} = 2 \sqrt{2 \ln(2)}\ \frac{\sigma_\mathrm{RMS}}{SR},
	\label{eq2}
\end{equation}

This equation details the electronic jitter component. $\sigma_{RMS}$ is the RMS of the noise value histogram of the amplification chain, $SR$ is the slew rate of the detection pulse and is defined as $SR = \mathrm{max}\left(\frac{\Delta V}{\Delta t}\right)$, where $\Delta V$ is the voltage difference of the pulse for the corresponding time difference $\Delta t$. $2\sqrt{2\ln{2}}$ is the RMS-to-FWHM factor. $\sigma_{RMS}$ has been measured by taking the RMS of the gaussian electronic noise distribution using a 6~GHz bandwidth oscilloscope. The SR was measured with the same oscilloscope. $\sigma_{RMS}$ typically equals 5.66~mV. $SR$ depends on the bias current and is typically in between 0.35~mV/ps and 1.3~mV/ps.\\

\subsection*{Setup jitter}

The setup jitter has been calculated from the following formula:
\begin{equation}
	j_\mathrm{setup} = \sqrt{j_\mathrm{pulse\ width}^2 + j_\mathrm{TCSPC}^2},
	\label{eq1}
\end{equation}

We experimentally measured $ j_\mathrm{TCSPC}$. The trigger pulse from the laser is divided in two pulses that are sent to the TCSPC module. We measured a gaussian distribution with a jitter (FWHM) of 9~ps. $j_\mathrm{pulse\ width}$ has been precisely characterized by the vendor (Nuphoton Technologies), it is a gaussian distribution with a pulse width (FWHM) of 6~ps. 

\section*{Superconducting molybdenum silicide film properties}

\begin{figure}[t!] 
	\includegraphics[width=160mm]{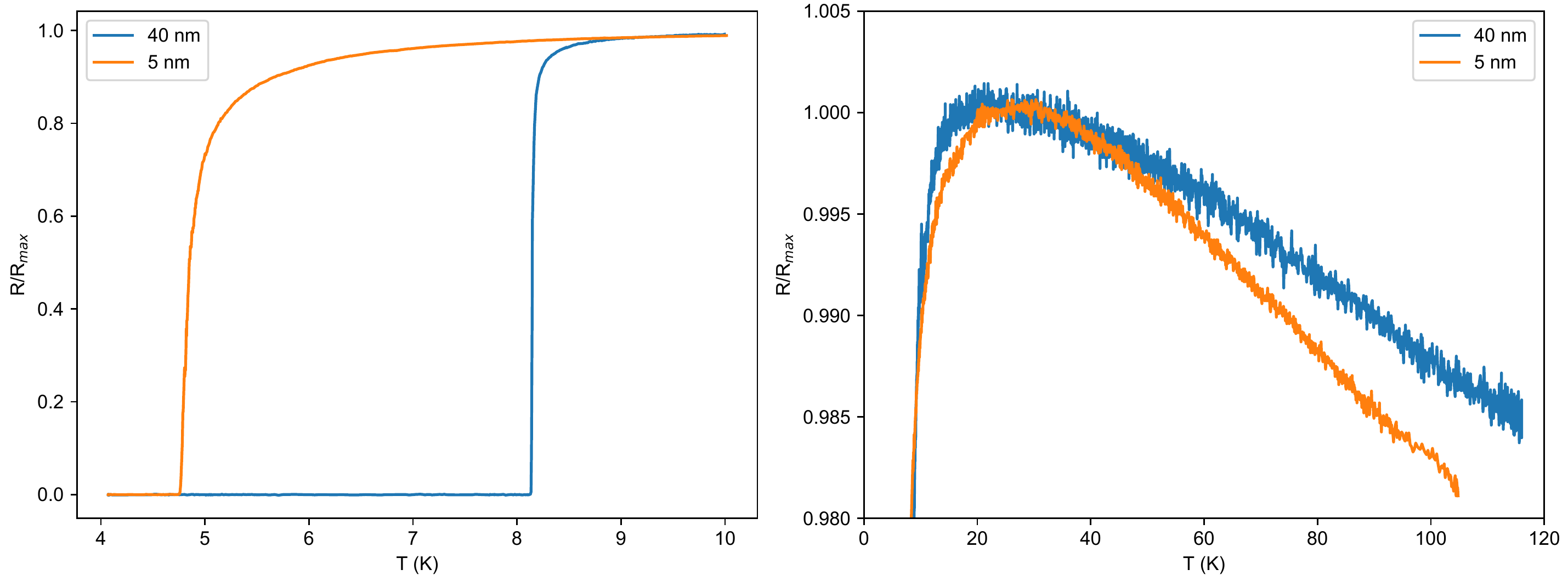}
	\caption{$R(T)$ measurement for two different thicknesses of Mo$_{0.8}$Si$_{0.2}$ deposited on silicium thermal oxide.}
	\label{fig3supp}
\end{figure}

\begin{figure}[t!] 
	\includegraphics[width=100mm]{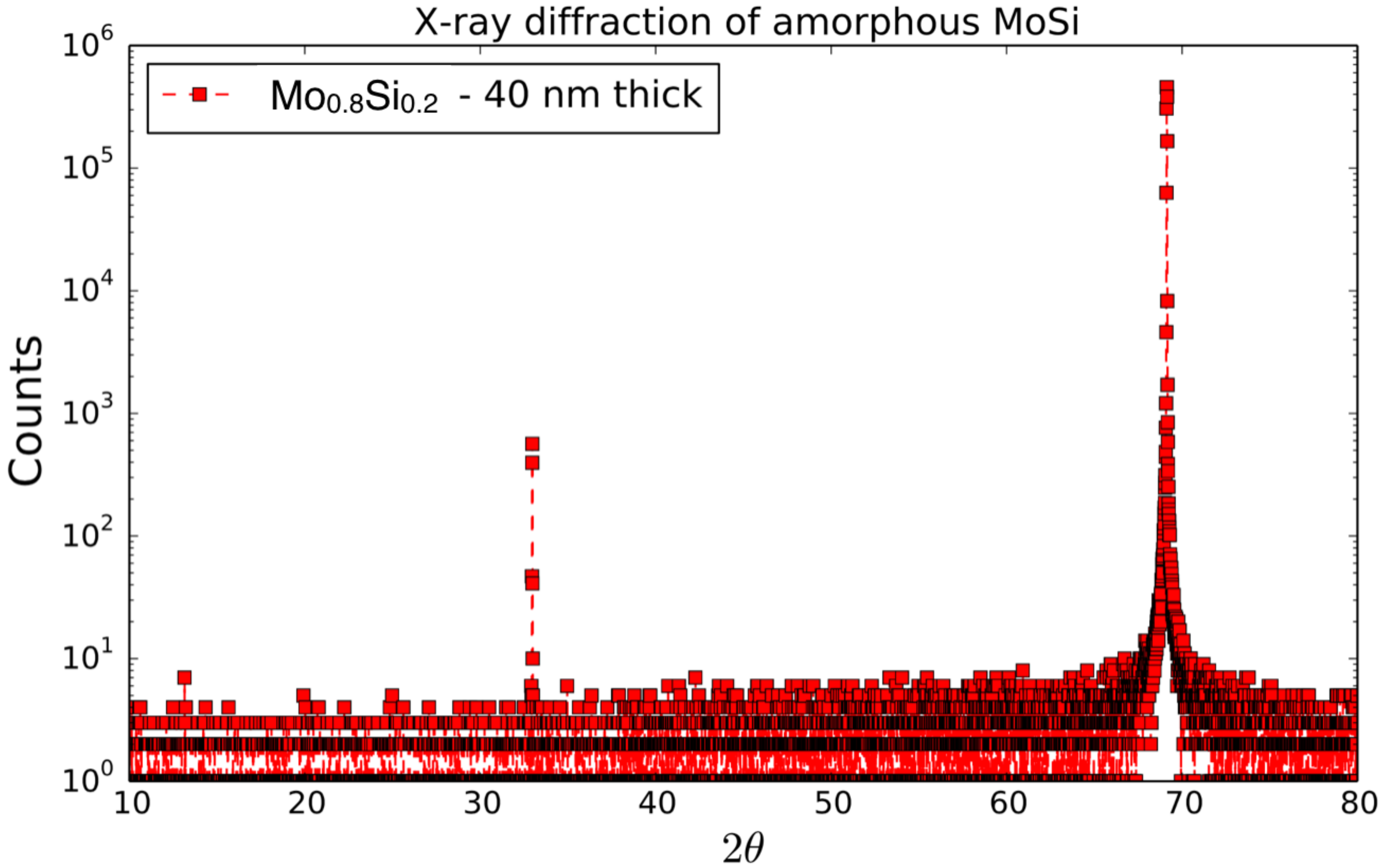}
	\caption{X-ray diffraction measurement of MoSi film deposited on silicium substrate. The two peaks are attributed to the silicium substrate.}
	\label{fig4supp}
\end{figure}

We measured the resistance $R$ as a function of the temperature $T$ for MoSi films with thicknesses of 5~nm and 40~nm, from 100~K down to 4.2~K and found a small, linear increase of $R$ for decreasing $T$ down to about 20~K. The Fig.~\ref{fig3supp} shows $R(T)$ for temperature close to $T_c$. This gives a residual-resistivity ratio (RRR) smaller than~1. For an unstructured 5~nm thick MoSi film, $R(300~K)/R(10~K)$ = 0.94. The increase was more pronounced for the thinner film, probably due to the importance of surface scattering. 
This ``disordered metal behaviour'' is commonly found in sputtered films, also of crystalline materials as NbN, when they are deposited under similar conditions. This is not specific to amorphous MoSi, and is therefore not sufficient to distinguish amorphous from crystalline materials. We performed X-ray diffraction measurement on MoSi films as shown in Fig.~\ref{fig4supp}, and these measurements were consistent with an amorphous film. The two observed peaks are attributed to the silicium substrate.\\

The room temperature resistance of our devices is a few $M\Omega$, depending on the geometry of the nanowire and of the meander. The current density at $I_{sat}$ is typically around 3~$MA/cm^2$ and is similar for all devices.\\

\end{document}